# CVTree for 16S rRNA: Constructing Taxonomy-Compatible All-Species Living Tree Effectively and Efficiently

Yi-Fei Lu (卢逸飞) [2], Xiao-Yang Zhi (职晓阳) [2,a,*] and Guang-Hong Zuo (左光宏)[1,b,*]

[1]*Wenzhou Institute, University of Chinese Academy of Sciences, Wenzhou, Zhejiang 325001, China*
[2]*Yunnan Institute of Microbiology, Key Laboratory of Microbial Diversity in Southwest China of Ministry of Education, School of Life Sciences, Yunnan University, Kunming 650091, China*

[*]Corresponding author E-mail: xyzhi@ynu.edu.cn (Zhi XY); ghzuo@ucas.ac.cn (Zuo GH)
[a] ORCID: 0000-0002-9862-377X.
[b] ORCID: 0000-0002-7822-5969.

**ABSTRACT**: The Composition Vector Tree (CVTree) method, developed under the leadership of Professor Hao Bailin, is an alignment-free algorithm for constructing phylogenetic trees. Although initially designed for studying prokaryotic evolution based on whole-genome, it has demonstrated broad applicability across diverse biological systems and gene sequences. In this study, we employed two methods, InterList and Hao, of CVTree to investigate the phylogeny and taxonomy of prokaryote based on the 16S rRNA sequences from All-Species Living Tree Project. We have established a comprehensive phylogenetic tree that incorporates the majority of species documented in human scientific knowledge and compared it with the taxonomy of prokaryotes. And the performance of CVTree were also compared with multiple sequence alignment-based approaches. Our results revealed that CVTree methods achieve computational speeds 1-3 orders of magnitude faster than conventional alignment methods while maintaining high consistency with established taxonomic relationships, even outperforming some multiple sequence alignment methods. These findings confirm CVTree's effectiveness and efficiency not only for whole-genome evolutionary studies but also for phylogenetic and taxonomic investigations based on genes.

# Introduction

Taxonomy serves as the foundational framework for deciphering biological diversity and evolutionary relationships, enabling systematic classification, comparative analyses, and evidence-based conservation strategies[1]. The advent of 16S rRNA gene sequencing revolutionized prokaryotic systematics by leveraging its dual characteristics: conserved regions that enable universal amplification and variable regions that provide phylogenetic resolution[2]. Prior to molecular techniques, taxonomic classification predominantly relied on error-prone phenotypic characterization. Carl Woese's seminal work in the 1970s established 16S rRNA as a molecular chronometer, reconstructing phylogenetic relationships and revealing the tripartite division of life into Bacteria, Archaea, and Eukarya[3]. This paradigm shifted not only redefined microbial phylogeny but also uncovered extensive archaeal diversity and the previously underestimated richness of uncultured environmental microbiota[4]. Subsequent technological advancements, including high-throughput sequencing platforms (e.g., Illumina[5]) and bioinformatics pipelines (RDP[6], Greengenes[7], SILVA[8], QIIME[9]), standardized 16S rRNA gene-based analyses, enabling large-scale microbiome studies and biogeographic investigations despite inherent limitations such as PCR amplification biases and restricted species-rank discrimination[10]. Contemporary initiatives like the All-Species Living Tree Project (LTP) further enhance taxonomic precision by integrating high-quality 16S rRNA sequences from validated type strains with genomic metrics to resolve phylogenetic inconsistencies[11].

Current phylogenetic reconstruction methodologies employing 16S rRNA gene sequences predominantly utilize multiple sequence alignment (MSA) tools, including CLUSTAL serial[12] (enhanced sensitivity through weighted sequence divergence), Muscle[13] (iterative refinement for high-throughput data), T-Coffee [14](hybrid global-local alignment), and MAFFT[15] (fast Fourier transform-based heuristics). While these algorithms preserve the positional homology essential for taxonomic assignment and support applications ranging from detecting horizontal gene transfer to identifying chimeras, they exhibit significant computational constraints. Progressive alignment strategies demonstrate poor scalability with increasing sequence numbers—MUSCLE encounters memory limitations beyond 5,000 sequences, while T-Coffee becomes computationally prohibitive. MAFFT exhibits reduced efficiency with datasets exceeding 10,000 sequences, and iterative refinement steps exacerbate resource requirements[16]. Despite emerging machine learning approaches and GPU-accelerated implementations, fundamental challenges remain in handling large-scale datasets characterized by extreme sequence divergence or intra-genomic heterogeneity, necessitating trade-offs between analytical precision and computational feasibility[17].

Alignment-free methodologies present a transformative approach to circumvent the computational bottlenecks inherent in traditional MSA-based phylogenetics[18]. The Composition Vector Tree (CVTree) algorithm, pioneered by Hao et al.[19], employs k-mer frequency analysis (fixed-length subsequences) to construct dissimilarity matrices directly from genomic sequences. This method has been successfully extended beyond its original application in bacterial and archaeal phylogeny [20–23] to encompass fungi[24,25], viral genomes [26], organellar DNA[27,28], and metagenomic datasets [29,30]. By eliminating alignment-dependent artifacts and genome size normalization requirements, CVTree achieves strain-rank discrimination validated against established taxonomic frameworks (e.g., Bergey's Manual)[31], demonstrating scalability that is compatible with large-scale initiatives like the LTP while providing complementary insights to marker gene-based approaches[32].

This study implemented the CVTree algorithm to reconstruct prokaryotic phylogenies using 16S rRNA sequences from type strains in the LTP database. We performed comparative analyses with conventional alignment-based methods through systematic benchmarking to evaluate computational efficiency and phylogenetic accuracy[33]. A comprehensive phylogenetic tree encompassing most documented prokaryotic species was constructed and systematically compared with established taxonomic classifications. Performance metrics revealed two key findings: First,

CVTree demonstrated 1-3 orders of magnitude faster computational speed compared to multiple sequence alignment (MSA)-based approaches. Second, employing an information entropy-based metric for quantifying topological congruence with taxonomic hierarchies, CVTree-derived phylogenies exhibited good consistency across multiple taxonomic ranks, matching and occasionally surpassing the performance of optimal MSA methods. These results validate CVTree's efficacy and efficiency for both whole-genome evolutionary studies and gene-based phylogenetic analyses. The integration of CVTree into 16S rRNA-based investigations offers novel perspectives for resolving prokaryotic evolutionary relationships and refining modern taxonomic frameworks.

## Materials and Methods

### The All-Species Living Tree Project

The All-Species Living Tree (LTP) project, initiated in 2007 through international collaboration, establishes a high-resolution phylogenetic framework for prokaryotic taxonomy by integrating 16S and 23S rRNA sequences from all validly published bacterial and archaeal type strains[34]. The framework utilizes curated entries from the SILVA database and maximum-likelihood algorithms to resolve taxonomic inconsistencies through dynamic phylogenetic trees that integrate phenotypic classification with molecular evolutionary relationships. The LTP database (release LTPs2024, used in this study) currently comprises 19,608 bacterial and 678 archaeal sequences, representing the most comprehensive reference for microbial diversity to date[35]. These sequences exhibit a mean length of 1461.9 bp (deviation=82.2). This standardized resource facilitates taxonomic validation, environmental sequence placement, and comparative genomics while maintaining backward compatibility with legacy rRNA studies. Recognized in major taxonomic compendia, the LTP exemplifies scalable phylogenetics that bridges classical microbiology with modern genomic approaches through reproducible, collaborative science.s

**Taxonomy of 16S rRNA**

Taxonomic classification is a dynamic system that undergoes continuous refinement as novel species are discovered, existing taxa are reorganized, and phylogenetic evidence is systematically integrated. While historical revisions of Bergey's Manual of Systematic Bacteriology [36] occurred periodically, digitalization and web-based platforms now facilitate rapid taxonomic updates, posing challenges for maintaining accurate species annotations. To address systematic gaps in species-rank classifications within the LTP dataset, we implemented two complementary strategies. For 16S rRNA sequences lacking species designations, species information was manually retrieved via NCBI Nucleotide database queries using unique GenBank accession numbers. Higher taxonomic ranks (genus and above) were reconstructed using genus-specific nomenclatural indicators embedded in species binomials[37], cross-referenced with the List of Prokaryotic names with Standing in Nomenclature (LPSN)—a curated repository reflecting current systematic consensus[38,39]. Targeted manual refinements reconciled discrepancies between phylogenetic relationships and traditional classifications. This integrated approach yielded fully validated taxonomic information for 19,508 bacterial and 677 archaeal strains (Data S1), ensuring synchronization with contemporary microbial systematics while preserving phylogenetic coherence.

**Phylogenetic Tree by CVTree**

The CVTree method employs an alignment-free phylogenetic analysis framework based on k-mer frequency statistics and Markov chain models. Genomic sequences are converted into high-dimensional vectors, with phylogenetic relationships quantified through cosine similarity or other distance metrics. Neighbor-joining algorithms then construct phylogenetic trees. Using the CVTree software suite (https://github.com/ghzuo/cvtree), we generated Newick-format trees through the InterList and Hao algorithms[40]. The detailed algorithms can refer to the manual of the CVTree software. For the Hao method, k-mer length was optimized using the criterion $log_m L < k < log_m L + 2$, with $m = 4$ represents genomic alphabet size, and L denotes the average sequence length. The reasonable $k$ for InterList methods in the CVTree software should be a little bigger than that of the Hao method and have a larger value range. A detailed discussion of this problem can refer to our previous works[33,41]. In this study, we set $k = 6$ for the Hao method and $k = 7$ for the InterList method.

**Phylogenetic Tree based on Alignment methods**

For comparison, we implemented a phylogenetic tree construction approach based on sequence alignment. Three widely used sequence alignment programs (MAFFT v7.525, Muscle v5.1, and Clustal Omega v1.2.4) were systematically employed to generate multiple sequence alignments (MSA) of 16S rRNA sequences derived from the LTPs project. And then phylogenetic trees were obtained by using FastTree v2.1.11[42]. Due to the extremely large dataset (over 20,000 gene sequences), we employed the Super5 algorithm, which is specifically designed for handling large-scale data in the Muscle software, and Clustal Omega (ClustalO)[43], which is the program for large datasets of CLASTAL serial. It should be noted that most MSA programs, including the three programs used in this study, leverage iterative convergence to achieve the optimal (self-consistent) alignment. Researchers can balance alignment accuracy against computational demands by controlling the number of iterations. In this study, default parameters of the programs, including three MSA programs and FastTree, were applied for all computations and comparative analyses to obtain a fair and reasonable environment for comparison. However, there is one exception when processing the whole LTPs2024 dataset with ClustalO. The default parameters failed to produce reasonable results (with $\widetilde{\Delta H} \sim 0.5$), therefore two additional iterations were used to obtain a reasonable result (in Table 1 and Figure 2).

**Evaluate Tree by Taxonomy**

Evaluating the quality of reconstructed phylogenetic trees is critical for assessing tree-building methodologies. While bootstrap values of individual branches or entire trees have been widely adopted as consistency metrics in numerous studies, our previous research demonstrated that integrating phylogenetic trees with independent biological classification systems provides a more objective evaluation criterion[33]. To implement this comparative analysis, we developed CLTree (available at https://github.com/ghzuo/collapse), a tool to rooting, annotate, and evaluate the phylogenetic tree by taxonomy[44]. It provides a way to root and evaluate the phylogenetic tree by a more independent information, i.e. the taxonomy system of the biology. In the algorithm, it identifies shared taxonomic features among terminal nodes within each branch and subsequently detects clades corresponding to specific taxonomic ranks. Notably, clade-based partitions at any taxonomic rank form non-overlapping divisions of species, mirroring the hierarchical segmentation in classification systems. In information theory, Shannon entropy ($H = \sum_i p_i \times log_2 p_i$) serves as an effective mathematical measure for quantifying discrepancies between these partitioning schemes. We mathematically demonstrate that clade-induced partitions constitute refinements of taxonomy-based divisions. Consequently, the ratio between the entropy reduction of clades relative to that of taxonomic systems, i.e. $\widetilde{\Delta H} = (H_{max} - H_{phy})/(H_{max} - H_{tax})$, provides an objective metric for evaluating phylogenetic-categorical congruence. Here $0 \leq \widetilde{\Delta H} \leq 1$, and $\widetilde{\Delta H} = 1$ means that all the taxa are monophyly on the phylogenetic tree. The operational protocols and theoretical foundations of CLTree, including detailed algorithmic implementations, are comprehensively documented in the software manual.

## Results and Discussion

**The All-Species Living Tree by CVTree**

Figure 1 showed a phylogenetic tree constructed by using CVTree based on 16S rRNA sequences from the All-Species Living Tree (LTPs2024) project[45] (Data S2). It was a challenge to display a phylogenetic tree with 20,286 leaves in a figure. Thus, we consolidated the branches sharing the same phylum into triangular nodes. The colors of the text and branches indicated that all sequences are distinctly segregated into two highest-rank clades in the phylogenetic tree, corresponding to the domains Bacteria and Archaea of the prokaryotes in taxonomy. That is, the phylogenetic relationships and taxonomy system are completely consistent at the domain rank.

At the phylum rank, all the 16S rRNA came from 42 phyla (4 archaeal and 38 bacterial). In the phylogenetic tree, they formed 61 phylum-rank clades. There were 36 phyla that were monophyletic, with only one clade in the phylogenetic tree, while the remaining six phyla were paraphyletic, forming 25 clades. Notably, the archaeal phylum *Methanobacteriota* was split into two segments due to the insertion of a monophyletic *Thermoproteota* lineage. Similarly, the phylum *Chloroflexota* (Bacteria) was divided into four branches, all resulting from a single insertion of the monophyletic phylum *Thermomicrobiota*. The phylum *Mycoplasmatota* was fragmented into three clades by two interspersed *Bacillota* lineages, with one clade containing an outlier strain. In contrast, the phylum *Actinomycetota* exhibited an inverse pattern: two strains were segregated into external clades. Although topological rearrangement could theoretically merge these clades, their significant topological divergence (coupled with truncated sequence lengths—less than one-third of the average length) suggests distinct evolutionary trajectories.

The paraphyletic phyla *Bacillota* and *Pseudomonadota* displayed greater complexity, each fragmented into seven dispersed clades. Phylogenetic tree topology adjustments failed to reconcile these clades into cohesive groups. Intriguingly, 15 *Pseudomonadota* and 8 *Bacillota* strains in the upper-right region of the tree were phylogenetically distant from their primary clades, potentially due to sequence truncation, which length are less than 1000 bp. However, the underlying causes of the remaining divergent branches may relate to the multiple origins and complex evolution of these groups themselves[46–48]. This analysis highlights the challenges in resolving monophyletic groupings for certain phyla using only the 16S rRNA sequences. The analysis based on the whole genomes may provide a more robust framework for understanding these taxonomic complexities.

**CVTree is Taxonomy-Compatible**

For comparative analysis, we evaluated multiple phylogenetic reconstruction methods using LTPs2024 16S rRNA sequences and quantified the number of monophyletic groups at each taxonomic hierarchy (Table 1). The results were systematically compared against phylogenetic references provided by the All-Species Living Tree Project (LTP), which employs sequence alignment-based tree construction using iterative refinements of 16S rRNA alignments from the SILVA database [11]. Notably, the SILVA repository contains 10-100×more sequences than those incorporated in the LTP project, potentially introducing confounding phylogenetic signals. To enable rigorous comparative analysis, we constructed the phylogenetic trees directly from the type strains 16S rRNA sequences from the LTPs project based on three MSA methods. Totally, there are six phylogenetic trees for the LTPs2024 16S rRNA sequences, i.e., two CVTree methods (InterList and Hao's method), FastTree implementations based on three MSA tools (Muscle, MAFFT, and ClustalO), and the reference tree provided by the LTP project. The six methodologies demonstrated comparable performance without statistically significant disparities. The Muscle-FastTree combination exhibited a slight advantage overall, but this refinement came at computational expense due to Muscle's intricate alignment process.

It should be noted that the 16S rRNA sequences in the LTP project were obtained through the collection of curated type strains, a process inherently subject to various technical and biological constraints. Consequently, the resulting dataset exhibits a highly uneven taxonomic distribution. At the phylum rank, for instance, the phylum *Pseudomonadota* is represented by 7,706 sequences, while six phyla each contain only a single sequence. Such extreme disparities in sequence representation underscore that reliance solely on quantitative counts of monophyletic groups would inevitably introduce systematic bias in phylogenetic analyses. Therefore, to more accurately assess the consistency between the phylogenetic tree and the classification system, we implemented the relative entropy reduction to describe the compatibility between phylogenetic tree and taxonomy (for computational methodology and definitions, refer to the Methods section).

Figure 2 presents a comparative analysis of the ratio of entropy reduction between phylogenetic trees generated through six distinct methodologies and the taxonomic system across multiple taxonomic ranks. In general, a progressive decline in entropy reduction ratios were observed among all methodologies as classification granularity increased from higher to lower taxonomic ranks. At the domain rank, complete entropy reduction ( $\widetilde{\Delta H} = 1$ ) was uniformly achieved by all six methods through the successful discrimination of two domains based on 16S rRNA profiles. However, none of the methodologies attained full resolution at subordinate classification ranks, with persistent paraphyletic groupings constituting a universal limitation. From phylum to genus ranks, the examined methodologies exhibited comparable performance characteristics, sustaining entropy reduction ratios exceeding 0.75. A marked performance degradation occurred at species-rank classification, where the mean ratio precipitously declined to approximately 0.6. This phenomenon may be attributed to dual factors: intrinsic limitations in 16S rRNA resolution capacity for species-rank differentiation[49–51], coupled with data instability arising from the LTP database's predominance of single-sequence reference strains that inherently preclude entropy reduction contributions.

Comparative analysis revealed methodological divergences: CVTree implementations demonstrated statistically superior performance at phylum-rank classification relative to other approaches, while ClustalO exhibited marginal advantages at class-rank categorization - a predictable outcome given its enhanced iterative refinement. Notwithstanding these localized variations, all six methodologies maintained equivalent overall performance metrics within this experimental framework, with CVTree variants showing context-specific optimization capacities in particular classification scenarios.

**Scaling Effect on Taxonomy-Compatible**

The results presented above were derived from a single analysis of all 16S rRNA sequences in the LTPs2024 project. To obtain statistically robust conclusions and investigate dataset size effects, we conducted systematic subsampling analyses. Specifically, subsets containing 1,000, 2,000, 4,000, 8,000, and 16,000 16S rRNA sequences were randomly drawn from the LTPs dataset, with ten independent replicates performed per sample size. As illustrated in Figure 3, phylogenetic trees were reconstructed using two CVTree-based algorithms and three MSA pipelines, followed by hierarchical quantification of relative entropy reduction against the established taxonomic system.

Given the limited discriminative power between the two domains (Bacteria and Archaea) and the prevalence of singleton species (represented by single sequences) at the species rank after subsampling, statistical analyses were restricted to five taxonomic ranks: phylum, class, order, family, and genus. Our findings reveal that all five methods exhibited data size-dependent variations in taxonomic compatibility, with more pronounced discrepancies observed at lower taxonomic ranks. Key observations include:

**Family/Genus ranks**: Performance degradation emerged at datasets ≥4,000 sequences.

**Class/Order ranks**: Efficacy decline became detectable only at the maximum tested size (16,000 sequences).

**Phylum rank**: Performance deterioration was exclusively observed in three MSA-based methods, while two CVTree-based methods kept their performance at higher ranks.

Overall, all five methods exhibit performance degradation as dataset size increases. This attenuation likely arises from increased experimental complexity due to an expanded data volume, potentially compounded by imbalanced dataset distributions. In comparative analysis: CVTree-based methods demonstrate relatively superior stability across heterogeneous datasets. The Muscle method performs comparably to CVTree, showing only marginal degradation at the phylum classification rank on the largest datasets. In contrast, MAFFT and ClustalO display notably inferior performance. Particularly for ClustalO, constrained by computational time limitations in our sampling framework preventing additional iterative refinement, its default parameter configuration demonstrates marked instability when handling datasets exceeding 8,000 sequences. The observed performance decline in MSA methods likely stems from accumulated errors during the alignment processes, which necessitated iterative correction. Notably, Muscle and MAFFT incorporate two iterative cycles by default, whereas ClustalO lacks default iteration. Nevertheless, CVTree maintains inherent algorithmic advantages with sequence-order-independent similarity calculations, thereby avoiding error accumulation and highlighting their enhanced reliability for constructing taxonomically congruent phylogenies in large-scale analyses. This systematic evaluation provides empirical evidence supporting CVTree-based methods as robust solutions for reconstructing evolutionary trees aligned with the taxonomic system.

**CVTree is much more Efficient**

Computational efficiency represents a critical evaluation metric in bioinformatics software. As illustrated in Figure 4, we present the average execution time (over ten replicates) required by five methods to construct phylogenetic trees from 16S rRNA sequences under varying dataset sizes. Among the three MSA tools, MAFFT demonstrates the fastest processing speed, outperforming the slowest method (Muscle) by approximately one order of magnitude. As mentioned above, MSA is an NP-hard problem, with computational demands escalating exponentially as dataset size increases, rendering it impractical for large-scale phylogenetic analyses. Existing heuristic MSA algorithms, such as MAFFT, Muscle, and ClustalO, prioritize either speed or accuracy but cannot guarantee globally optimal alignments. These observations underscore the persistent trade-off between precision and efficiency in MSA workflows.

The two CVTree-based methods show slight differences in computational time between them, yet both achieve remarkable speed advantages over MSA-based approaches, surpassing even the fastest MSA method (MAFFT) by over an order of magnitude. Notably, CVTree was originally designed for whole-genome analysis, where its performance benefits from shorter sequence lengths while maintaining the capability to enhance taxonomic resolution through sequence elongation - particularly effective for species- and subspecies-rank discrimination. These characteristics collectively establish CVTree as a sequence-based method with a strong potential for taxonomic identification.

Figure 4 also reveals an important trend: although CVTree maintains a two-order-of-magnitude advantage overall, the computational time differences between methods gradually diminish with increasing sequence counts. This convergence stems from theoretical computational complexity considerations - when excluding additional iterations in MSA pipelines, all five methods encounter the same algorithmic bottleneck during the tree reconstruction. Both FastTree and CVTree employ the neighbor-joining method [52] in their workflows, which carries an $O(n^3)$ time complexity. Thus, the advantage of CVTree may be kept for datasets containing less than millions of sequences, further scaling beyond this range would necessitate either novel algorithmic developments or substantial workflow optimizations to address inherent computational constraints.

## Conclusion

In this study, we conducted a systematic comparison of two CVTree-based alignment-free methods and three heuristic MSA algorithms (MAFFT, Muscle, and ClustalO) using the LTPs 2024 16S rRNA dataset, evaluating their accuracy, robustness, and computational efficiency. Among MSA tools, MAFFT demonstrated unparalleled processing speed, significantly outperforming other algorithms. However, its accuracy exhibited a pronounced decline with increasing dataset sizes. In contrast, Muscle achieved superior accuracy and stability, albeit at the cost of computational resources, which scaled quadratically to cubically with sequence volume. Notably, heuristic MSA algorithms often incorporate iterative refinement options. Our analysis of ClustalO revealed that default parameter configurations (e.g., no iterations) led to reduced accuracy on large datasets, whereas increasing iteration counts improved performance. These observations underscore the persistent trade-off between precision and efficiency in MSA workflows.

To overcome the limitations imposed by this "accuracy-efficiency trade-off," methodological innovation is critical. Alignment-free approaches, exemplified by CVTree, present a compelling alternative for phylogenetic reconstruction and taxonomic classification. Originally developed for whole-genome evolutionary analysis, CVTree has demonstrated outstanding computational efficiency, surpassing the fastest MSA tool, MAFFT, by orders of magnitude faster. In terms of accuracy, CVTree matched the performance of the top-tier MSA method, Muscle, and exceeded it in some metrics. These results position CVTree as a robust, scalable, and efficient algorithm for prokaryotic phylogeny and classification using 16S rRNA data, advocating its broader application in single-gene-based taxonomic studies beyond traditional whole-genome frameworks.

## Author Contributions

GZ designed the study, GZ and YFL performed the analysis, and GZ and XYZ wrote the manuscript.

## Competing interests

The authors have declared that no competing interests exist.

## Acknowledgments

GHZ thanks the Wenzhou Institute, University of Chinese Academy of Sciences (Grant No. WIUCASQD2021042).

**Table 1** The number of monophyletic taxon units with multiple genomes on the phylogenetic trees built by different methods. The number in parentheses in the first column indicates the number of taxonomic units that contain multiple genomes and unique genome.

| Taxon Rank | InterList | Hao | ClustalO | Muscle | MAFFT | LTPs |
|---|---|---|---|---|---|---|
| Domain (2+0) | 2 | 2 | 2 | 2 | 2 | 2 |
| Phylum (36+6) | 30 | 27 | 30 | 30 | 24 | 27 |
| Class (77+23) | 58 | 54 | 57 | 58 | 52 | 54 |
| Order (184+54) | 120 | 112 | 115 | 121 | 117 | 117 |
| Family (472+132) | 269 | 264 | 268 | 287 | 277 | 282 |
| Genus (1950+1726) | 1325 | 1287 | 1224 | 1323 | 1269 | 1204 |
| Species (208+19694) | 94 | 97 | 100 | 111 | 104 | 115 |

# Figure Legends

**Figure 1 The all-species living tree built by CVTree.**

In this phylogenetic tree, branches belonging to the same phylum are collapsed into a triangle and are color-coded according to their phylum, corresponding to the surrounding color bars. The upper-level branches and phylum names are marked in blue or red, depending on whether they belong to Bacteria or Archaea, respectively.

**Figure 2 The ratio of entropy reduction at different taxonomic ranks.**

Six phylogenetic trees, constructed based on all LTPs2024 16S rRNA sequences, were each compared with the taxonomic system and labeled with different symbols and colors. These include two CVTree algorithms, InterList and Hao; as well as trees built using three multiple sequence alignment software, Muscle, MAFFT, and ClustalO, followed by the FastTree method; and the phylogenetic tree downloaded directly from the LTPs 2024 project.

**Figure 3 Boxplot for the ratio of entropy reduction as a function of the size of the dataset for different taxonomic ranks**

For datasets of varying sizes, the statistical performance at different taxonomic ranks was evaluated using the CVTree method (InterList and Hao) and FastTree construction based on three multiple sequence alignment methods (Muscle, MAFFT, and ClustalO). Different colors were used to distinguish the various methodological workflows.

**Figure 4 Time cost as a function of the size of the dataset for five methods.**

The computation time mentioned here refers to the time required to generate a phylogenetic tree from gene sequences, tested on a 40-core CPU with 32 threads, and averaged with variance calculated based on 10 repeated runs. Different colors and symbols in the figure are used to mark the various software methodological workflows.

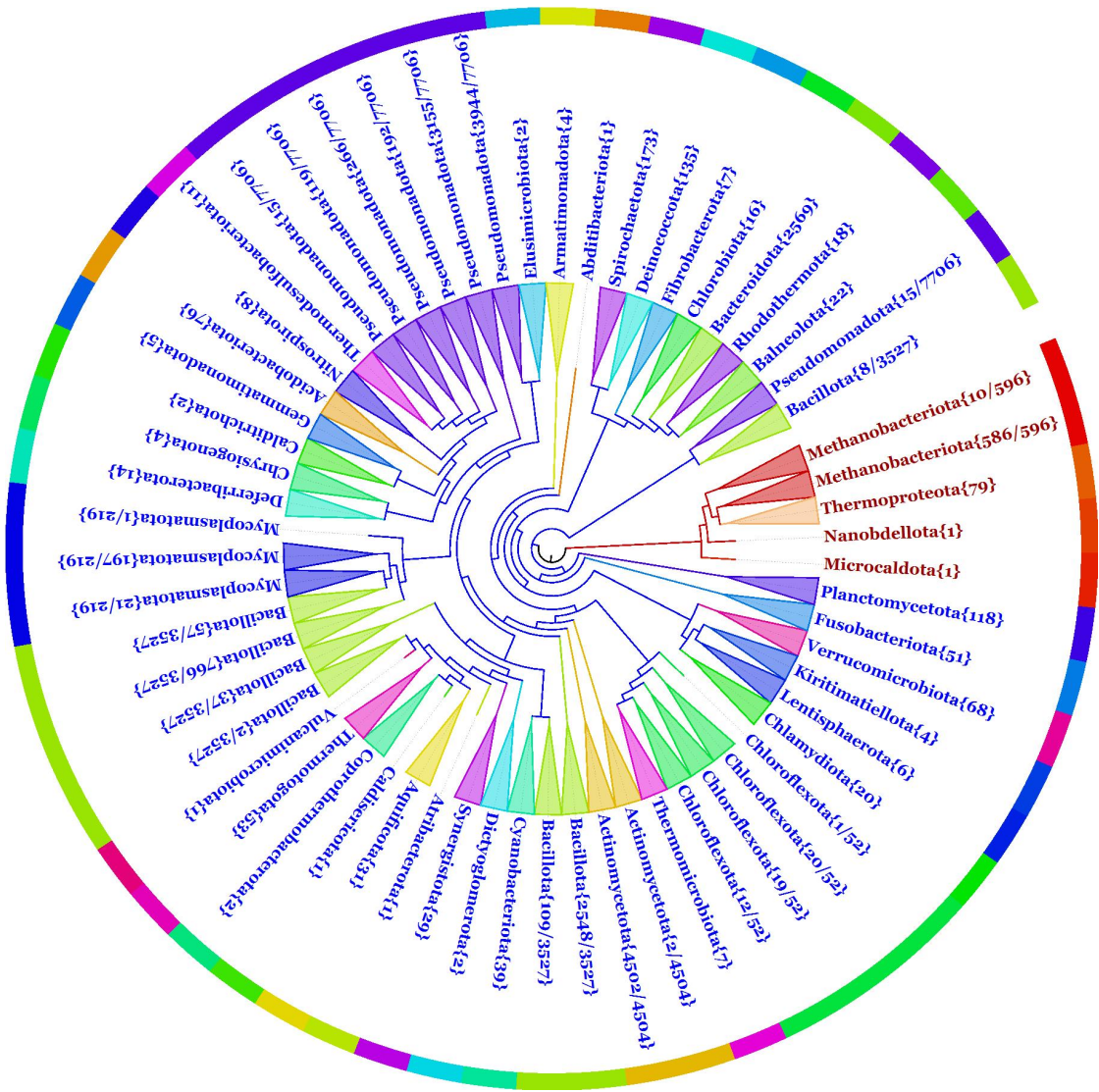

**Figure 1 The all-species living tree built by CVTree.**

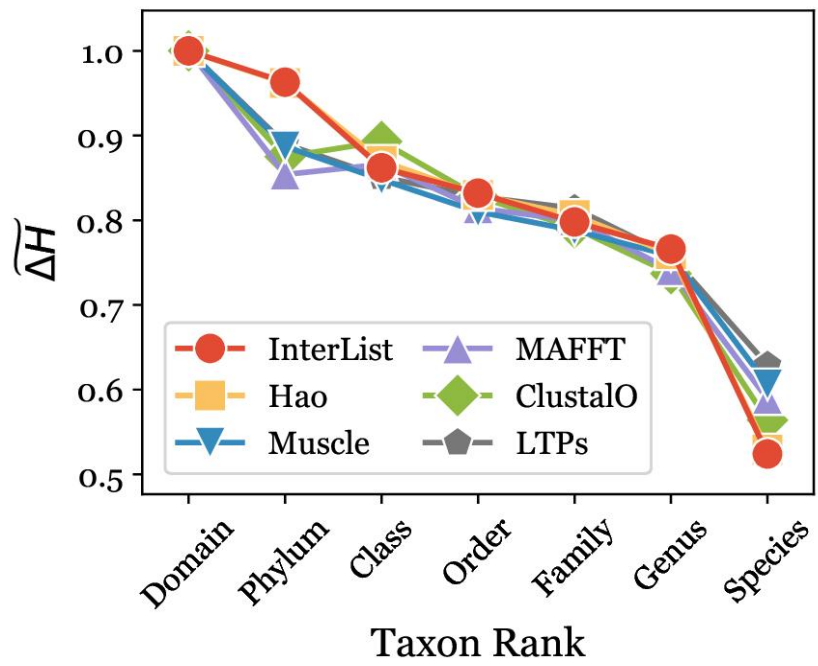

**Figure 2 The ratio of entropy reduction at different taxonomic ranks.**

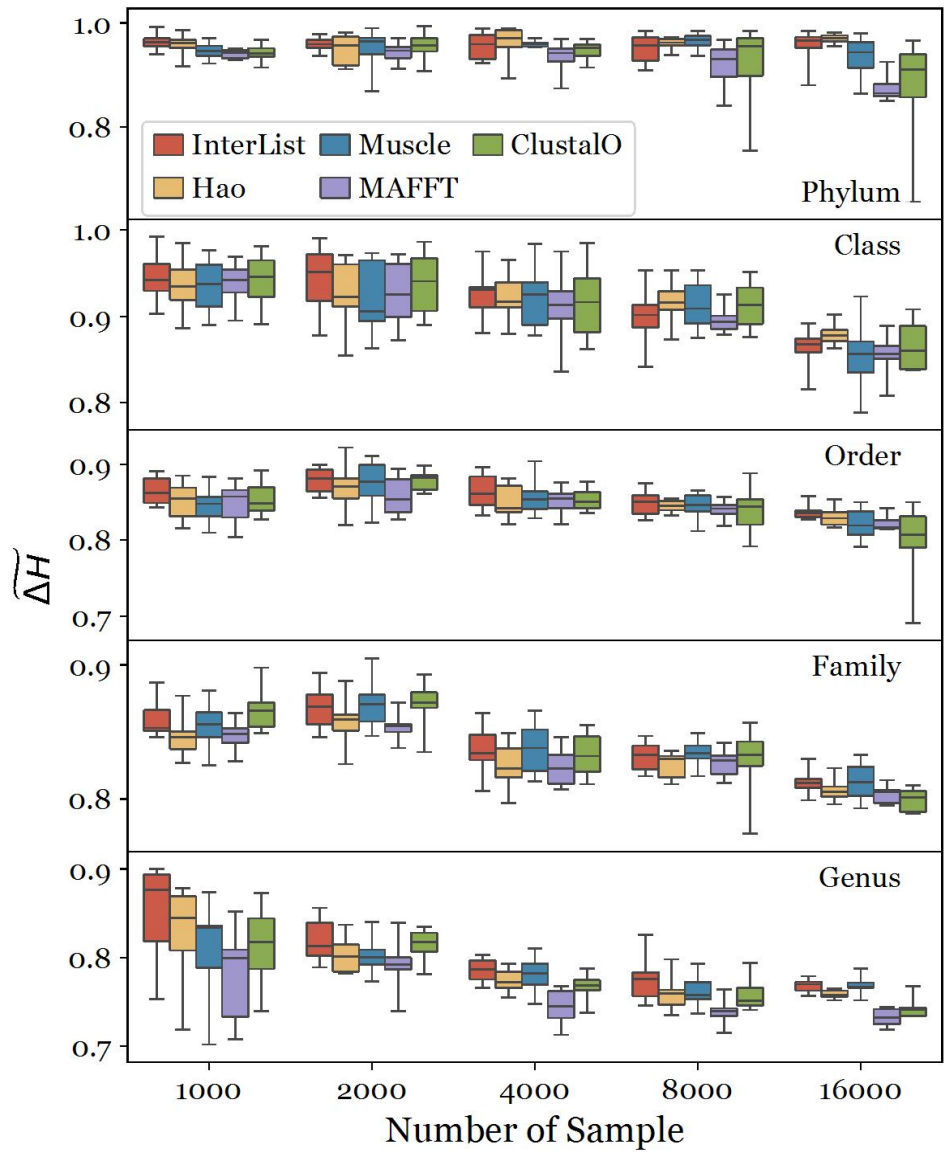

**Figure 3 Boxplot for the ratio of entropy reduction as a function of the size of the dataset for different taxonomic ranks.**

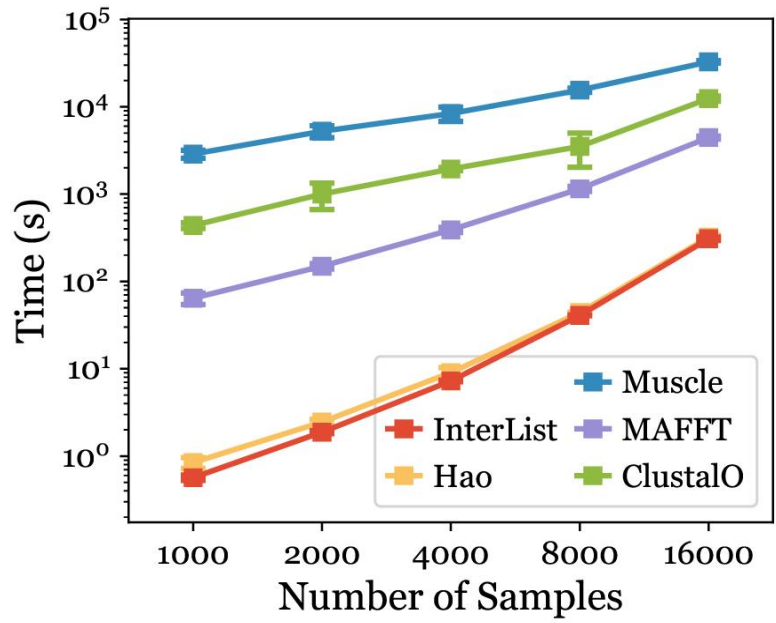

**Figure 4 Time cost as a function of the size of the dataset for five methods.**